\documentclass[manuscript]{acmart}
\setcopyright{none}

\usepackage{algorithmic}
\usepackage{algorithm}
\usepackage{listings}
\usepackage{subcaption}
\usepackage[utf8]{inputenc}
\usepackage{makecell}
\usepackage{float}
\usepackage{tabularx}
\usepackage{graphicx}
\usepackage{caption}

\AtBeginDocument{%
  \providecommand\BibTeX{{%
    \normalfont B\kern-0.5em{\scshape i\kern-0.25em b}\kern-0.8em\TeX}}}

\acmPrice{15.00}
\acmISBN{978-1-4503-XXXX-X/18/06}

\begin{document}

\title{NFTs to MARS: Multi-Attention Recommender System for NFTs}
%(v2)\title{NFT Recommendations with Multi-Modal Multi-Objective and Attention Mechanism}


\author{Seonmi Kim}
\authornote{These authors contributed equally to this research.}
\email{msraask3@unist.ac.kr}
\orcid{0009-0003-3566-6312}
\author{Youngbin Lee}
\authornotemark[1]
\email{young@unist.ac.kr}
\author{Yejin Kim}
\authornotemark[1]
\email{kimyejin99@unist.ac.kr}
\author{Joohwan Hong}
\authornote{Corresponding author.}
\email{joohwanhong@unist.ac.kr}
\author{Yongjae Lee}
\authornotemark[2]
\email{yongjaelee@unist.ac.kr}
\affiliation{%
  \institution{} \\ 
  \institution{Ulsan National Institute of Science and Technology}
  \streetaddress{50, UNIST-gil}
  \city{Ulsan}
  \country{South Korea}
  \postcode{44919}
}

\begin{abstract}
%(v1)
Recommender systems have become essential tools for enhancing user experiences across various domains. While extensive research has been conducted on recommender systems for movies, music, and e-commerce, the rapidly growing and economically significant Non-Fungible Token (NFT) market remains underexplored. The unique characteristics and increasing prominence of the NFT market highlight the importance of developing tailored recommender systems to cater to its specific needs and unlock its full potential. In this paper, we examine the distinctive characteristics of NFTs and propose the first recommender system specifically designed to address NFT market challenges. 
% We have compiled a comprehensive dataset that includes all available NFT transactions for four popular collections from the blockchain, providing a solid foundation for our study. 
In specific, we develop a Multi-Attention Recommender System for NFTs (\texttt{NFT-MARS}) with three key characteristics: (1) graph attention to handle sparse user-item interactions, (2) multi-modal attention to incorporate feature preference of users, and (3) multi-task learning to consider the dual nature of NFTs as both artwork and financial assets. We demonstrate the effectiveness of NFT-MARS compared to various baseline models using the actual transaction data of NFTs collected directly from blockchain for four of the most popular NFT collections. The source code and data are available at \url{https://anonymous.4open.science/r/RecSys2023-93ED}.
\end{abstract}

% <concept_id>10002951</concept_id>
% <concept_desc>Information systems</concept_desc>
% <concept_id>10002951.10003317.10003347.10003350</concept_id>
% <concept_desc>Information systems~Recommender systems</concept_desc>
% Multimedia and multimodal retrieval; • Computing methodologies
% → Multi-task learning

% \ccsdesc[500]{Information systems~Recommender systems}
% \ccsdesc[300]{Multimedia and multimodal retrieval}
% %\ccsdesc{}
% \ccsdesc[500]{Computing methodologies~Multi-task learning}

%\keywords{Non Fungible Tokens, Recommender Systems, Multi-Task Learning, Multi-Modal Learning, Graph Attention Network}



\maketitle

% Adjust these values to change the spacing around subsubsections

\setlength{\textfloatsep}{0.3cm}
\section{Introduction}

Where cryptocurrency was proposed as the digital substitute for
currency, NFTs are nowadays being touted as the digital substitute for collectibles. With NFTs being a growing trend, the total worth of NFT market reached \$41 million in 2021 alone, nearing that of the entire global fine art market \cite{Conti2022nftexp}. 
It was only at the beginning of 2020, when the NFT market started to take off \cite{vairagade2022proposal}. 
There were 1,415,638 NFT sales with a trading volume of 82,492,916 USD in 2020, which later soared to 27,414,477 sales and a trading volume of 17,694,851,721 USD in 2021 \cite{besancia2022nftreport}. 
Among them, most of the transactions identified are for collectibles which take up to 27\% of the total sales. Given the significant growth and unique characteristics of the NFT market, it is crucial to understand its potential in today's digital landscape. This creates opportunities to leverage recommender systems for maximizing benefits from digitized transaction information.

However, NFT recommender systems face three unique challenges. First, \textit{extremely sparse user-item interactions.} NFTs are absolutely lacking in feedback information since NFT’s uniqueness allows only one person to own it at a time while music, movies, and products on online platforms can be consumed by multiple users simultaneously and therefore are more likely to have plenty amount of feedback information. Second, \textit{anonymity}. It is difficult to utilize various user features since NFTs operate in a blockchain system, which is inherently anonymous. Third, \textit{dual nature of NFTs as both artistic and financial assets}. Hence, users in the NFT market have diverse purchasing motivations, which can range from buying NFTs as pure investments to buying them as a form of digital collectibles. Therefore, a recommender system addressing these aspects is essential to tackle these challenges.

In this paper, we develop an NFT recommender system, called \textbf{Multi-Attention Recommender System for NFTs (\texttt{NFT-MARS})}, to address these challenges. It consists of three key components: (1) graph attention to handle sparse user-item interactions, (2) multi-modal attention to incorporate user-specific feature preferences, and (3) multi-task learning to address the dual nature of NFTs as artworks and investment assets. We have collected actual transaction data directly from blockchain networks for four of the most popular NFT collections, and our experimental results show significant improvement in recommendation performance for NFT collectibles. This is the first study to propose a recommender system specifically designed for NFT collectibles and demonstrate its performance using real-world data.

 %which provides a unified way to utilize multi-modal features (images, text data, and numeric data) and effectively focus on the most relevant parts of the features to make decisions. 
%(v1) In this paper, we try to tackle aforementioned challenges by proposing a novel graph attention network-based model. Our model comprises of three key components : (1) multi-modal graph-based representation learning phase, which constructs parallel graphs between user and item nodes, similar to previous approaches (cite MGAT); (2) cross attention mechanism to identify the varying importance of different features by aggregating information from multiple graphs; and (3) multi-objective model training to predict both relevant items for each user and the price movement (binary value) of the corresponding item. Our experimental results confirm our expectations, with our model significantly improving recommendation performance by effectively incorporating all types of item features and outperforming other models.

\section{Related Work}

\subsection{Graph-Based Recommender Systems.} 
Recent advancements in graph-based recommender systems have been successful in addressing user-item interaction sparsity. Graph Convolutional Networks (GCN) \cite{kipf2016semi} and Graph Attention Networks (GAT) \cite{velickovic2017graph}, are two such approaches, with GAT further enhancing the approach by incorporating attention mechanisms. GAT’s multi-modal learning combines information from various sources by constructing parallel user-item graphs to improve representation learning \cite{baltescu2022itemsage, tao2020mgat, wei2019mmgcn}. In addition to multi-modal learning, GAT have also been enhanced by integrating multi-task learning. This allows them to perform multiple recommendation-related tasks simultaneously within a single model, further improving their efficacy \cite{huang2021graph, li2020multi, hsu2021fingat}. Our approach extends GAT by incorporating multi-modal, multi-attention, and multi-task learning to address the unique challenges associated with NFT data.

\subsection{NFT Recommender Systems.} 
The idea of applying a recommender system to the NFT sector was first proposed in a blog article, where a multiple regression-based recommender system was proposed \cite{opensea2020}. This system utilized previous purchase patterns and NFTs in users' wallets to estimate the likelihood of similar wallets owning NFTs from specific categories. However, this method primarily focuses on broad categories and necessitates manual input from users. Moreover, it is sensitive to data sparsity. Subsequently, more recent research has proposed content-based recommendation aproaches \cite{piyadigama2022analysis,piyadigama2022exploration}. These methods incorporate the similarity scores or rarity scores of items, aiming to tailor recommendations based on these features. The focus on item features sometimes overlooks the importance of adapting to the varying preferences and interest of the users. Therefore, there is room for further development in creating recommender systems that balance item attributes with user preferences in the NFT market. To address the limitations of previous studies, we design a graph-based reocmmender system to fully leverage high-order connections between entities using both user purchase record and item features.

\subsection{Recommender Systems in Related Fields.}
NFTs share similarities with traditional art, secondhand, and stock markets. First, as digital artworks,  NFTs possess both aesthetic and investment value derived from content characteristics, similar to traditional art markets \cite{frye2021nfts}. Their unique ownership aspect results in sparse user-item feedback, resembling challenges in art and secondhand markets. Artwork recommender systems utilize content-based approaches, considering various features like metadata and visual elements to offer personalized suggestions amid sparse user-item interactions \cite{messina2019content, semeraro2012folksonomy}. Second, NFT markets are similar to secondhand markets in a sense that an item can be possessed by a single user in most cases. Recommender systems for secondhand markets also favor content-based techniques over collaborative filtering, as sold items lack purchase records \cite{yu2020personalized}. To mitigate the cold-start problem, MultiRec \cite{rashed2020multirec} employs multi-task learning, while \shortcite{wang2021personalized} proposes models leveraging multi-modal inputs.
Third, some users view NFTs as investment assets like stocks. Stock recommender systems aim to improve future price forecasting accuracy while considering user preferences, often using graph-based methods \cite{hsu2021fingat, ying2020time}. FinGAT \cite{hsu2021fingat}, for example, identifies profitability and user interests through multi-modal data and multi-task learning. Drawing on these insights, we use a multi-attention structure to handle both the sparsity and multi-modal feature of NFTs and employ multi-task learning to handle the dual nature of NFTs as artworks and investment assets.
  
\section{Data}\label{sec:Data}
\subsection{Source and Description}
In this study, we focus on the  four top-ranked ERC-721-based \cite{ERC721} NFT collectibles, which are Bored Apes Yacht Club (BAYC), Cool Cats, Doodles, and Meebits, in terms of market capitalization as of August 2022 according to \cite{dappradar2022}. We collected content features from the OpenSea API  \cite{openseaapi2022} and transaction data from Etherscan \cite{etherscan2022} and web3.py, covering the entire period between September 1\textsuperscript{st} 2021 and March 10\textsuperscript{th}\ 2023. Our dataset consists of implicit feedback and considers purchase history as user-item interactions. To mitigate the sparsity of the NFT transactions, we implemented a minimum threshold of five interactions for users in our user-item matrices, excluding users with fewer than five interactions, and incorporated item and user features to enhance the predictive power. The data includes 4,575 users, 25,014 items, and 56,981 interactions across all four collections. For detailed descriptions and further analysis of the data used, please refer to Appendix \ref{sec: Data Description}. The data is available at \url{https://anonymous.4open.science/r/RecSys2023-93ED}.

\subsection{NFT collection characteristics}
Many NFT collections feature a large number of NFTs that share a similar artistic style. One prominent example is the Bored Ape Yacht Club (BAYC), which comprises of 10,000 NFT artworks depicting variants of Bored Apes. Each token in BAYC shares same array of traits including ‘fur’, and ‘clothes’, depicting the color and style of its fur and outfit, respectively, with slight variations across each token. The rarer the features of each token, the higher the value it is likely to fetch. For example, a trait like ‘dark brown fur’ could be found in 14\% of the collection, making corresponding tokens less rare, whereas a trait like ‘solid gold fur’, found in only 0.46\%, adds uniqueness and rarity to the token in terms of aesthetic value.

However, as NFTs function as both artwork and financial investment, it is crucial to consider their aesthetic and financial aspects. Sections \ref{sec:3.3} and \ref{sec:3.4} will discuss item and user features impacting recommendations, and furthermore, the most influential factors that affect recommendations across collections will be examined in Section \ref{sec:5.2}.

\subsection{Item Features} \label{section_3_3}
\label{sec:3.3}
NFT item features, including content and transaction attributes, provide insights into aesthetic and financial value, assisting in matching user purchasing behavior. The item features we used are as follows:

\begin{itemize}
\item\textbf{\textit{Image}}:
The most critical feature for understanding the visual aspects of NFTs, as digital artworks, is the image. We downloaded images from the URL of each token using OpenSea API. 

\item\textbf{\textit{Text}}:
The text feature offers insights into rarity factors and complements the image by capturing potentially overlooked visual elements. We sourced tag data from the OpenSea API, which includes tags characterizing each NFT's visual properties, such as "background color," "body," and "outfit."

\item \textbf{\textit{Price}}:
The average selling price in Ethereum (ETH) and Wrapped Ethereum (WETH) is considered. Prices quoted in other currencies like USDC and DAI are replaced with 0 due to valuation issues.

\item\textbf{\textit{Transaction}}:
The frequency of transactions associated with a particular item can be indicative of its popularity. We used the average holding period for all transactions of an item.
\end{itemize}

\subsection{User Features}
\label{sec:3.4}
Utilizing user features in NFT recommendations provides insights into individual purchasing behaviors and preferences, enabling more accurate suggestions that are aligned with the user's price range and investment strategy. The list of features we used includes:

\begin{itemize}
\item\textbf{\textit{Price}}: The average purchase price of each user was used to represent the user's financial capability and willingness to pay for NFTs, calculated in the same manner as the price feature of items.

\item\textbf{\textit{Transaction frequency}}: The user's average holding period for purchased tokens can provide insight into the user's trading behavior. This feature was calculated in the same manner as the transaction feature of items.

\item\textbf{\textit{Transaction count}}: To gauge the level of activity of a user in the market, we calculate the total number of transactions associated with each individual wallet address as a representative metric.
\end{itemize}

\section{Method}
\subsection{Problem Definition}

\begin{figure*}[!htbp]
\centerline{\includegraphics[width=1\columnwidth]{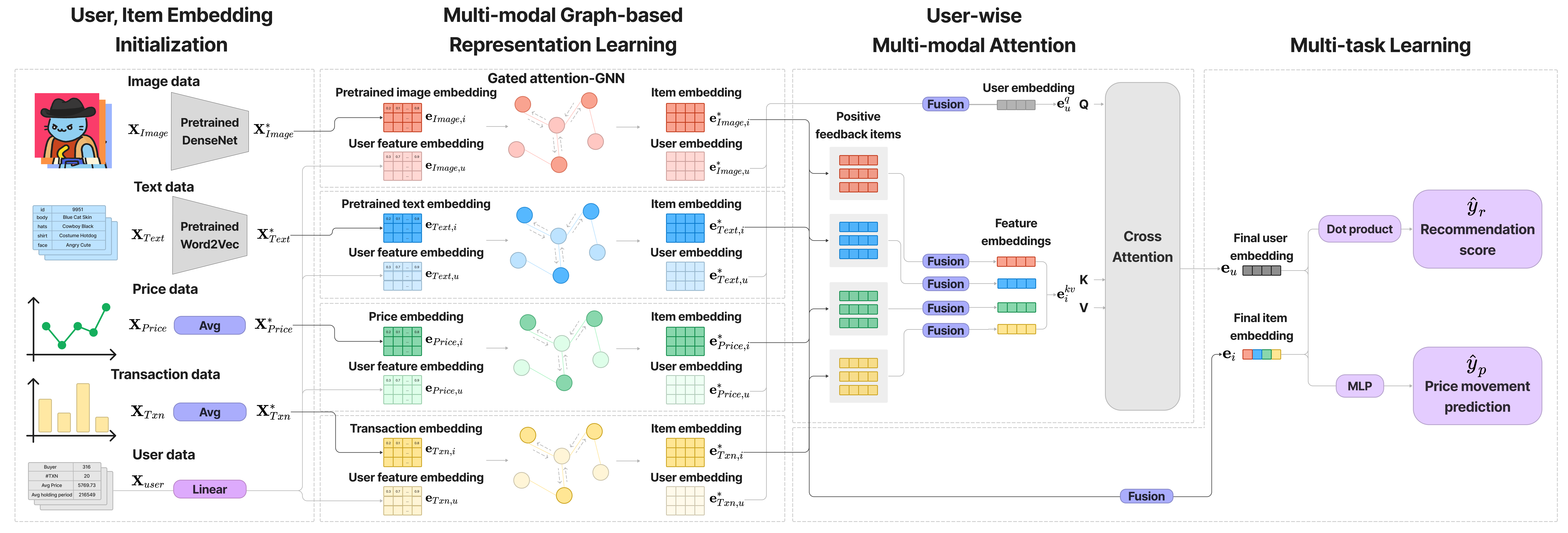}}
\setlength{\abovecaptionskip}{2pt}
\captionsetup{skip=2pt}
\caption{Model architecture of \texttt{NFT-MARS}}
\label{model architecture}
\end{figure*}

We consider each wallet address as a unique user $u$, and each token as each item $i$, with user set $U$ and item set $I$. The user-item interaction matrix $\mathbf{Y} \in \mathbb{R}^{|U| \times |I|}$ contains implicit feedback, where $y_{ui}=1$ if a user $u$ bought item $i$, and 0 otherwise. From $\mathbf{Y}$, we build multi-modal bipartite graphs $\mathcal{G}_{m}$ where modality indicator $m$ belonging to a set of modalities $\mathcal{M} = $\{Image, Text, Price, Transaction\}.
We also have a price movement label matrix, $\mathbf{P}$ for multi-task learning, where $p_{ui}$ is assigned a value of 1 if the price of item $i$ increases in subsequent transaction, and 0 otherwise. Our goal is to predict a user's interest in item $i$, learning a prediction function $\hat {y}_{ui}= \mathcal F(u,i|\Theta, \mathbf{Y}, \mathcal{G}_{m}, \mathbf{P})$, where $\hat {y}_{ui}$ denotes the probability that user $u$ buys item $i$ and $\Theta$ denotes the trainable model parameters of function $\mathcal F$.

The model framework begins with preparing various input features for multi-modal graphs (Section \ref{section_4_2}). These features are used to initialize node embeddings of multi-modal graphs, which are refined by propagation according to the graph structure (Section \ref{section_4_3}). User embeddings are updated using attention mechanism to reflect feature preferences of users (Section \ref{section_4_4}). Lastly, multi-task learning incorporates the dual nature of NFTs as artworks and financial assets to the final recommendation task (Section \ref{section_4_5}).

\subsection{Features Preparation} \label{section_4_2}

\textbf{Item.} We create a representation of each raw item feature, utilizing a total of four item features within their respective modality graphs and get $\mathbf{X^*}_{m} = {f_m}\left(\mathbf{X}_{m}\right)$ from each modality $m$, where $f$ differs depending on the modality of the inputs. We use pretrained DenseNet\cite{iandola2014densenet} and Word2Vec\cite{mikolov2013distributed} models for extracting features from image and text data, respectively. The average price and transaction features are duplicated to create 64-dimensional vectors. For more detailed explanations on the preparation of side information, please refer to Appendix \ref{sec: Side Information Preparation}.

\noindent
\textbf{User.}
We create a user feature matrix $\mathbf{X}_{user}$ that consists of the user features discussed in Section \ref{section_3_3}.

\begin{figure}[!hbtp]
\begin{minipage}[t]{0.5\textwidth}%
\begin{algorithm}[H]
\caption{Graph attention for each modality}
    \label{alg:1}
    \begin{algorithmic}
        \REQUIRE User-item interaction Matrix $\mathbf{Y}$, Multi-modal bipartite graphs $\mathcal{G}_{m}$, Price movement label matrix $\mathbf{P}$, User features $\mathbf{X}_\text{user}$, Item features $\mathbf{X}^*_m$
        \ENSURE Node representations $\mathbf{e}_{m,h}^*$
        \FORALL{$u \in U, m \in \mathcal{M}$}
            \STATE $\mathbf{e}_{m,u} \gets \textit{tanh} ( \mathbf{W}_{m,u} \cdot \mathbf{X}_\text{user}) $
        \ENDFOR
        \FORALL{$i \in I, m \in \mathcal{M}$}
            \STATE $\mathbf{e}_{m,i} \gets \mathbf{X}^*_m$
        \ENDFOR
        \FORALL{$h \in U \cup I, m \in \mathcal{M}, l \in L$}
            \STATE $\mathbf{e}^{(l)}_{m,N_h} \gets \textit{LReLU}(\sum_{t \in N_h} f_a(h, t) \cdot f_g(h, t) \cdot \mathbf{W}_{m,1} \cdot \mathbf{e}_{m,t})$
            \STATE $\tilde{\mathbf{e}}^{(l)}_{m,h} \gets \textit{LReLU}(\mathbf{W}_{m,2} \cdot \mathbf{e}_{m,h} + \mathbf{e}_h)$
            \STATE $\mathbf{e}^{(l)}_{m,h} \gets \textit{LReLU}(\mathbf{W}_{m,3} \cdot \mathbf{e}_{m,N_h} + \tilde{\mathbf{e}}_{m,h})$
        \ENDFOR
        \STATE $\mathbf{e}_{m,h}^* = \textit{concat}_{l}(\mathbf{e}^{(l)}_{m,h})$
    \end{algorithmic}
\end{algorithm}
\end{minipage}%
\hfill
\begin{minipage}[t]{0.48\textwidth}%
\begin{algorithm}[H]
\caption{User-wise multi-modal attention}
    \label{alg:2}
    \begin{algorithmic}
        \REQUIRE Multi-modal graph node representations $\textbf{e}_{m,h}^*$
        \ENSURE User embedding $\textbf{e}_u$, Attention score $\textbf{s}_u$
        \FORALL{$u \in U$, $m \in \mathcal{M}$}
            \STATE $\textbf{e}_u^q \leftarrow \frac{1}{|\mathcal{M}|} \sum_{m} \textbf{e}_{m,u}^*$
            \STATE $\textbf{Q} \leftarrow \textbf{W}_q \cdot \textbf{e}_u^q $
            % \STATE $\textbf{e}_{u}^{kv} \leftarrow \textit{concat}_{m}(\frac{1}{|Pos_u|} \sum_{i \in Pos_u} \textbf{e}_{m,i}^*)$
            \STATE
            $\textbf{e}_{u}^{kv} \leftarrow \textit{concat}_{m}(\frac{1}{|Pos_b|} \sum_{i \in Pos_b} \textbf{e}_{m,i}^*)$       
            \STATE $\textbf{K} \leftarrow \textbf{W}_{k} \cdot \textbf{e}_{u}^{kv}$
            \STATE $\textbf{V} \leftarrow \textbf{W}_{v} \cdot \textbf{e}_{u}^{kv}$\\
            \STATE $\textbf{e}_u \leftarrow \textit{softmax}\left(\frac{\textbf{Q} \textbf{K}^{\top}}{\sqrt{d_k}}\right) \textbf{V}$
            \STATE $\textbf{s}_u \leftarrow \frac{\textbf{Q} \textbf{K}^{\top}}{\sqrt{d_k}}$
        \ENDFOR
    \end{algorithmic}
\end{algorithm}
\end{minipage}%
\end{figure}

\subsection{Graph Attention for Each Modality} \label{section_4_3}

\textbf{Node Embedding Initialization.} Here, we initialize node embeddings, which are input to the graphs. For modality $m$, we make both user embeddings and item embeddings as $d_{m}$-dimensional vectors. Specifically, as described in Algorithm \ref{alg:1}, the user embedding $\mathbf{e}_{m,u} \in \mathbb{R}^{d_m}$ is obtained from passing the user feature matrix $\mathbf{X}_{user}$ through a linear layer, and the item embedding $\mathbf{e}_{m,i} \in \mathbb{R}^{d_m}$ is obtained from the item feature representation in Section \ref{section_4_2}. We also create the $d$-dimensional ID embedding for each node $h$, $\mathbf{e}_{h}$, which is randomly initialized and optimized during training to project representations from different modalities into the same dimension.

\noindent
\textbf{Embedding Propagation Layer.}
Next, we use a gated attention GNN to refine the user and item embeddings, effectively utilizing sparse interactions. Specifically, 
for each ego node $h$, we perform three steps. (1) We aggregate messages $\mathbf{e}_{m, N_h}$, where $N_h$ is the set of indices of its neighboring nodes, using propagation gates $f_g(h, t)$ and attention score $f_a(h, t)$ using the weight matrix $\mathbf{W}_{m,1}$, as described in \cite{tao2020mgat}, which is learned in this step.
(2) We perform a linear transformation of the node embedding $\mathbf{e}_{m,h}$ using the transformation matrix, $\mathbf{W}_{m,2}$, and its corresponding id embedding, $\mathbf{e}_{h}$, to project representations from different modalities into the same dimension $d$.
(3) The aggregated information $\mathbf{e}_{m, N_h}$ and projected embedding $\tilde{\mathbf{e}}_{m, h}$ are then combined using a transformation and addition operation, where $\mathbf{W}_{m,3}$ denotes the trainable weight matrix, to perform cross-modality propagation and produce the final node representation $\mathbf{e}_{m, h}^{*}$ of the graph. The propagation process can be expanded to multiple hops by adding extra $L$ embedding propagation layers.

\subsection{User-wise Multi-modal Attention} \label{section_4_4}

This section delineates the enhancement of user embeddings through the incorporation of a cross-attention mechanism, which catpures the relative importance of each modality for a specific user. Initially, updated user embeddings are generated by combining embeddings from all four graphs, which are then utilized to derive queries ($\mathbf{Q}$). This involves consolidating user embeddings from various modalities, represented as $m \in \mathcal{M}$, into a single user embedding, denoted as  $\mathbf{e}_u^q$, following the procedure outlined in Algorithm \ref{alg:2}.

Similarly, we create feature embeddings by merging the item embeddings for each modality, which are then used to get keys and values ($\mathbf{K}$, $\mathbf{V}$). To understand the significance of each feature, we focus solely on positive items within a batch. We represent the set of items with positive feedback by users in a batch as $Pos_b$ and create a fusion of item embeddings from $Pos_b$ by taking average for each modality.

During cross-attention, the neural network learns the cross-attention weights $\mathbf{W}_q$, $\mathbf{W}_k$, and $\mathbf{W}_v$ by processing the user embedding $\mathbf{e}_u^q$ and the feature embedding $\mathbf{e}_u^{kv}$, with each represented as a $d_k$ dimensional vector. These weights, combined with the attention computation, allow the model to focus on the most important information for each user by weighing the different modalities accordingly. The final user embedding after the multi-modal attention is denoted as $\mathbf{e}_u$ and the user's attention scores for different features are also obtained as $\mathbf{s}_{u}$.

\subsection{Multi-task Learning} \label{section_4_5}

\textbf{Task 1: Recommendation Score.}
Item embeddings are combined using the multi-modal representation learning method, calculated as $
\mathbf{e}_{i} = {1 \over |M|} \sum_{m \in M}\mathbf{e}_{m, i}^{*}$. The recommendation score function, $\hat{y}_{r} = \mathbf{e}_{u}^{T}\mathbf{e}_{i}$, predicts items a user is likely to purchase by computing the inner product of the final user embedding $\mathbf{e}_{u}$ and the item embedding $\mathbf{e}_{i}$. Given the scores $\hat{y}_{r}$, the recommendation loss function is defined as:
\begin{equation} \label{eu_eqn}
\mathcal{L}_{r} = \sum_{(u,i,j) \in O} -ln  \sigma (\hat{y}_{r}-y_{r})
\end{equation}
where $O \in \{(u,i,j)|(u,i) \in R^{+}, (u,j) \in R^{-}\}$ present the training dataset, $R^{+}$ denotes the positive (observed) interaction between user $u$ and item $i$, while $R^{-}$ is the negative (unobserved) interaction between user $u$ and item $j$; $\sigma$ is the sigmoid function. We use BPR loss to optimize recommendation scoring functions.

\noindent
\textbf{Task 2: Price movement prediction.}
To predict whether an item's price will rise or fall, we use the price movement prediction function,
$
\hat{y_{p}} = MLP(\mathbf{e}_{u} \lVert \mathbf{e}_{i})
$, where $\mathbf{e}_{i}$ is the item embedding described in the recommendation score. The user and item embeddings are concatenated and passed through an MLP layer. Given the score $\hat{y_{p}}$, we define price movement loss function as follows:
\begin{equation} \label{eu_eqn}
\mathcal{L}_{p} = - \sum_{(u,i) \in O}y_{p} \cdot log(\hat{y}_{p})+(1-y_{p}) \cdot log(1-\hat{y}_{p})
\end{equation}
\noindent
where $O$ denotes the training dataset and positive (observed) interaction between user $u$ and item $i$, same as the above-mentioned notations. We use binary-cross entropy loss, which is commonly used to optimize price movement prediction signals effectively.

\noindent
\textbf{Optimization.}
We employ an end-to-end learning approach to optimize the \texttt{NFT-MARS} model, using multiple objective functions. The overall objective function, $\mathcal{L}(\theta)$, is minimized by calculating the weighted sum of loss functions, with $\alpha$ representing the model's hyperparameter, as follows:
\begin{equation} \label{eu_eqn}
\mathcal{L} = (1-\alpha) \cdot \mathcal{L}_{r}+\alpha \cdot \mathcal{L}_{p}
\end{equation}

\section{Experiment}
\label{sec:Experiment}

\subsection{Experimental Settings}
\textbf{Data sets.} 
We conducted experiments on four real-world NFT transaction datasets BAYC, Cool Cats, Doodles, and Meebits (See Section \ref{sec:Data} for details.). In order to ensure an adequate size for the test set, we randomly selected 40\% of the interactions from each user to be included in the test set. Consequently, the historical interactions between users and items within each collection were, on average, divided into a 62\% training set and a 38\% testing set. The testing set was further split equally into validation and test subsets through random selection. We performed negative sampling to create user-specific pairwise preferences, designating user-purchased items as positive and selecting five negative samples for each positive sample based on popularity.

\noindent
\textbf{Evaluation schema.} 
We used NDCG@K (Normalized Discounted Cumulative Gain) and Recall@K to evaluate the performance of the top $K$ recommendations. Evaluation was carried out on positive items from the test set along with 100 negative items sampled based on item popularity. 
 
 % We randomly split the historical user-item interaction of each collection into training, validation and test sets with proportions of $80\%$, $10\%$, and $10\%$, respectively. Moreover, we evaluate the model on the set of positive items in testing set with their sampled negative items. Per each positive item, 100 negative items are sampled based on item popularity. 
 
\noindent
\textbf{Model comparison.}
Our proposed \texttt{NFT-MARS} model uniquely combines a graph-based collaborative filtering (CF) approach with a content-based method, incorporating both user and item features. We compared it against various CF, content-based, and graph-based models, including Pop, ItemKNN \cite{deshpande2004item}, BPR \cite{rendle2012bpr}, DMF \cite{xue2017deep}, NeuMF \cite{he2017neural}, LightGCN \cite{he2020lightgcn}, FM \cite{rendle2010factorization}, DeepFM \cite{guo2017deepfm}, WideDeep \cite{cheng2016wide}, DCN \cite{wang2017deep}, AutoInt \cite{song2019autoint}, and MGAT \cite{tao2020mgat} to evaluate the effectiveness of our model. Pop is a simple popularity-based recommendation approach, while ItemKNN is a model-based method that uses item similarities. BPR, and DMF are matrix factorization models optimized using Bayesian Personalized Ranking, and deep matrix factorization, respectively. NeuMF combines the linearity of matrix factorization with the non-linearity of neural networks for enhanced recommendation performance. FM, DeepFM, and WideDeep are feature-based models employing factorization machines, deep neural networks, or a combination of both. DCN is a deep and cross network model, AutoInt is an auto-encoder based interaction model, while LightGCN and MGAT are graph-based models utilizing graph convolutional networks and multi-modal graph attention networks, respectively. The experiments for baseline models were conducted using RecBole \cite{zhao2022recbole}. Each experiment was carried out three times, with the average score being recorded in Table \ref{table_1}.

\noindent
\textbf{Hyperparameter settings.} We optimized our model using the Adam optimizer, with a learning rate of 0.01, and 50 epochs. Hyperparameter tuning involved a random search using Recall@50 as an indicator, with search ranges including the dimensions ($d$) of the graph's final node representation [128, 512], loss alpha ($\alpha$) [0.1, 0.2], batch size [1024, 4096], number of hops ($L$) [1, 2, 3], and regularization weight [0.1, 0.001]. Baseline models' hyperparameters were also tuned, in regards to embedding size, learning rate, and dropout ratio. 
As each NFT collection has unique characteristics, separate models were built and fitted for each collection, resulting in different best hyperparameters for each NFT collection. Further information regarding hyperparameters is presented in Appendix \ref{sec: Hyperparameter}.

\noindent
The code and more detailed experiment settings are in \url{https://anonymous.4open.science/r/RecSys2023-93ED}.

\begin{table*}[t]
\scriptsize{
\captionsetup{font=tiny}
\fontsize{5.6}{8}\selectfont
  \captionsetup{skip=2pt}
\caption{Comparison with baseline models}
\label{table_1}
\begin{tabular*}{\linewidth}{@{\extracolsep{\fill}}lrrrrrrrrrrrrrr}

\toprule
         &            & Pop    & ItemKNN         & BPR    & DMF    & NeuMF  & LightGCN        & FM              & DeepFM & WideDeep & DCN             & AutoInt & MGAT            & \textbf{\texttt{NFT-MARS}}            \\
Dataset  & Metric     &        &                 &        &        &        &                 &                 &        &          &                 &         &                 &                 \\ \hline
\midrule
BAYC     & Recall@30 & 0.1782 & 0.2010          & 0.1747 & 0.1972 & 0.1496 & 0.1520          & 0.1307          & 0.1571 & 0.1853   & 0.1947          & 0.1672  & \textbf{0.2412} & \textbf{0.3153} \\
         & NDCG@30   & 0.0605 & \textbf{0.0780} & 0.0639 & 0.0668 & 0.0465 & 0.0615          & 0.0448          & 0.0529 & 0.0574   & 0.0665          & 0.0518  & 0.0714          & \textbf{0.0945} \\ \cline{3-15} 
         & Recall@50 & 0.3031 & 0.3346          & 0.3136 & 0.3280 & 0.2724 & 0.3025          & 0.2505          & 0.2857 & 0.3273   & 0.3260          & 0.3153  & \textbf{0.4516} & \textbf{0.5211} \\
         & NDCG@50   & 0.0866 & 0.1056          & 0.0924 & 0.0942 & 0.0718 & 0.0913          & 0.0689          & 0.0793 & 0.0867   & 0.0939          & 0.0821  & \textbf{0.1104} & \textbf{0.1329} \\ \hline
Cool Cats & Recall@30 & 0.1987 & 0.2234          & 0.2114 & 0.2012 & 0.1517 & 0.1983          & 0.1797          & 0.1869 & 0.1778   & 0.1982          & 0.1536  & \textbf{0.2681} & \textbf{0.3527} \\
         & NDCG@30   & 0.0666 & 0.1151          & 0.1054 & 0.0679 & 0.0601 & \textbf{0.1058} & 0.0767          & 0.0605 & 0.0560   & 0.0705          & 0.0467  & 0.0802          & \textbf{0.1056} \\ \cline{3-15} 
         & Recall@50 & 0.3274 & 0.3472          & 0.3312 & 0.3359 & 0.2685 & 0.3380          & 0.3004          & 0.3147 & 0.3096   & 0.3273          & 0.3020  & \textbf{0.4596} & \textbf{0.5379} \\
         & NDCG@50   & 0.0930 & \textbf{0.1405} & 0.1299 & 0.0959 & 0.0840 & 0.1336          & 0.1013          & 0.0870 & 0.0833   & 0.0975          & 0.0769  & 0.1163          & \textbf{0.1403} \\ \hline
Doodles  & Recall@30 & 0.1838 & 0.2384          & 0.1978 & 0.2136 & 0.1515 & 0.1538          & 0.1317          & 0.1556 & 0.1938   & 0.2587          & 0.1737  & \textbf{0.2807} & \textbf{0.2837} \\
         & NDCG@30   & 0.0593 & \textbf{0.0954} & 0.0762 & 0.0724 & 0.0476 & 0.0674          & 0.0484          & 0.0460 & 0.0612   & \textbf{0.0902} & 0.0539  & 0.0807          & 0.0798          \\ \cline{3-15} 
         & Recall@50 & 0.3202 & 0.3666          & 0.3531 & 0.3641 & 0.2816 & 0.3261          & 0.2600          & 0.3022 & 0.3476   & 0.4122          & 0.3313  & \textbf{0.4519} & \textbf{0.5167} \\
         & NDCG@50   & 0.0867 & \textbf{0.1216} & 0.1073 & 0.1030 & 0.0739 & 0.1007          & 0.0736          & 0.0753 & 0.0923   & \textbf{0.1216} & 0.0854  & 0.1128          & \textbf{0.1231} \\ \hline
Meebits  & Recall@30 & 0.1926 & 0.3287          & 0.3459 & 0.2743 & 0.3477 & 0.3569          & 0.3566          & 0.3922 & 0.3857   & 0.4096          & 0.3779  & \textbf{0.6189} & \textbf{0.6461} \\
         & NDCG@30   & 0.0559 & 0.1586          & 0.2474 & 0.0918 & 0.2501 & \textbf{0.2653} & \textbf{0.2531} & 0.2557 & 0.2101   & 0.2495          & 0.2287  & 0.1970          & 0.2274          \\ \cline{3-15} 
         & Recall@50 & 0.3955 & 0.5032          & 0.5333 & 0.4629 & 0.4808 & 0.5717          & 0.5634          & 0.5593 & 0.5618   & 0.5578          & 0.5416  & \textbf{0.7627} & \textbf{0.7891} \\
         & NDCG@50   & 0.0972 & 0.1949          & 0.2843 & 0.1309 & 0.2773 & \textbf{0.3076} & \textbf{0.2939} & 0.2900 & 0.2464   & 0.2803          & 0.2618  & 0.2241          & 0.2541          \\ \hline
\bottomrule

\end{tabular*}
}
\end{table*}

\subsection{Results and Discussion}
\label{sec:5.2}

\textbf{RQ1. Performance comparison with the baselines.} 
Our model (\texttt{NFT-MARS}) outperformed the baseline models by a significant margin, as shown in Table \ref{table_1}, with higher average Recall@30,50 and NDCG@30,50 scores across all datasets. This suggests that utilizing graph-based recommender systems with additional item features effectively models the intricate relationship between users and items and utilizes neighborhood information.

\begin{table}[t]
\scriptsize{
\captionsetup{font=tiny}
\captionsetup{skip=2pt}
\caption{Ablation Studies}
\label{table_2}
\fontsize{5.6}{8}\selectfont
 \begin{tabular*}{\linewidth}{@{\extracolsep{1.5pt plus 1pt minus 1pt}}l*{17}{c}}
    \toprule
    \multicolumn{7}{c}{\textbf{Study 1: Performance by varying $\alpha$}} & \multicolumn{1}{c}{} & \multicolumn{8}{c}{\textbf{Study 2: Comparison with single-graph models}} \\
    \cmidrule(lr){1-7} \cmidrule(lr){9-16}
  &    $\alpha$     &    0.0 &   0.1 &   \textbf{0.2} &   0.3 &   0.4   & &  & Input & Image & Text & Price & Txn & All & \textbf{\texttt{NFT-MARS}} \\ 
Dataset  & Metric     &        &                 &                 &        &          &  & Dataset     & Metric&           &         &                 &           &    &      \\ \hline
\midrule
BAYC & Recall@50 & 0.5487 & 0.5662 & \textbf{0.5998} & 0.5737 & 0.5600 & & BAYC & Recall@50 & 0.5424 & 0.4988 & 0.4428 & 0.4780 & \textbf{0.5560} & 0.5332 \\
        & NDCG@50 & 0.1638  & 0.1669  & \textbf{0.1815}  & 0.1548  & 0.1582 & & & NDCG@50 & 0.1424 & 0.1334 & 0.1167 & 0.1177 & \textbf{0.1464} & 0.1302 \\ \cmidrule(lr){1-7} \cmidrule(lr){9-16}
Cool Cats & Recall@50 & 0.5690  & 0.5822  & \textbf{0.5994}  & 0.5951 & 0.5691 & & Cool Cats & Recall@50 & 0.4014 & 0.5151 & 0.3839 & 0.4153 & 0.3782 & \textbf{0.5258} \\
        & NDCG@50 & 0.1601 & \textbf{0.1755} & 0.1609 & 0.1541  & 0.1633 & & & NDCG@50 &0.1068 & 0.1337 & 0.0890 & 0.1012 & 0.0876 & \textbf{0.1407} \\ \cmidrule(lr){1-7} \cmidrule(lr){9-16}
Doodles & Recall@50 & 0.5419 & \textbf{0.5814}  & 0.5544  & 0.5636  & 0.5710 & & Doodles & Recall@50 & 0.4475 & 0.4798 & 0.4288 & 0.4742 & 0.4254 & \textbf{0.5083}\\
        & NDCG@50 & 0.1352 & 0.1429 & 0.1379  & \textbf{0.1576} & 0.1502 & & & NDCG@50 & \textbf{0.1286} & 0.1217 & 0.1237 & 0.1141 & 0.1040 & 0.1179 \\ \cmidrule(lr){1-7} \cmidrule(lr){9-16}
Meebits & Recall@50 & 0.7278  & 0.7272  & 0.6557  & 0.7298  & \textbf{0.7597} & & Meebits & Recall@50 & 0.6864 & 0.7817 & 0.7344 & 0.7214 & 0.6707 & \textbf{0.8220} \\
        & NDCG@50 & \textbf{0.2833} & 0.2220  & 0.2100 & 0.2298  & 0.2526 & & &NDCG@50 & 0.2284 & 0.2277 & 0.1746 & 0.2105 & 0.2461 & \textbf{0.2570}\\

\bottomrule
\end{tabular*}
}
\end{table}

\begin{figure}[t]
  \begin{minipage}{\textwidth}
    \centering    \includegraphics[width=1.0\linewidth]{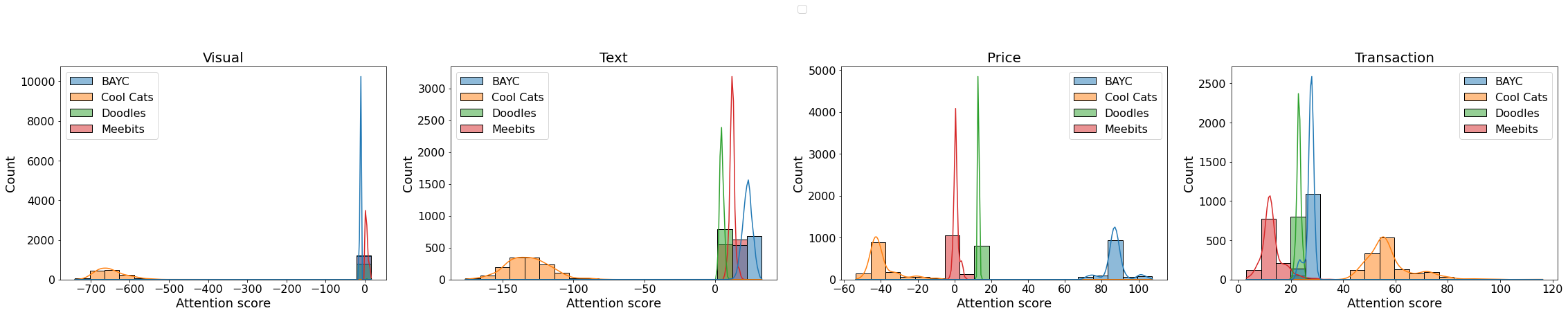}
    \captionsetup{skip=2pt}
    \caption{Attention score distributions in different collections}
    \label{Figure_2}
  \end{minipage}
\end{figure}

\noindent
\textbf{RQ2. Ablation studies.}
To further assess the effectiveness of our architectural choices, we conducted ablation studies.

$-$ \textbf{\textit{Study 1: Effectiveness of multi-task learning.}}
First, we evaluated the impact of balancing NFTs as artwork and financial assets in our model by varying $\alpha$ from 0 to 0.4. The results in Table \ref{table_2} showed that models with non-zero $\alpha$ values outperformed single-task learning models. This indicates that considering the financial aspects of NFTs in the recommendation process can enhance the overall user experience. %within the NFT market. 

$-$ \textbf{\textit{Study 2: Effectiveness of multi-modality.}}
We compared the performance of our multi-modal model against single-graph models with single modality (Image, Text, Price, Transaction (Txn)) as well as another single-graph model that simply concatenates all modalities (All). As shown in Table \ref{table_2}, our multi-modal model outperformed most of the single-graph models in terms of Recall and NDCG values. Even in BAYC, we can see that the performance of \texttt{NFT-MARS} is quite close to the best among single-graph models. This suggests that it is important to effectively incorporate multiple modalities in NFT recommender system.

\noindent
\textbf{RQ3. Multi-modal attention score analysis.}
In Figure \ref{Figure_2}, it is interesting to note that the attention score distribution for various features is significantly different across each collection. For Cool Cats, all features except for transaction showed negative values, indicating that user preferences were dominated by transaction frequency for items. In the case of BAYC, price was the most important feature. For Doodles, both transaction and price were the two influential factors.  Meebits leans towards transactions as the leading feature, although not as dominantly as in Cool Cats. A common trend across all four collections is the notably low attention scores for the image feature, suggesting that the visual attributes of NFTs may not be as pivotal in driving user preferences. This could be attributed to the fact that critical visual attributes are often represented in text descriptions. These varying attention scores across NFT collections underline the importance of considering user-specific multi-modal attention.

\section{conclusion}
In this paper, we proposed the first recommender system specifically designed to address various challenges posed by the unique characteristics of NFTs. Our model \texttt{NFT-MARS} was designed to (1) address extreme sparsity of user-item interactions, (2) incorporate multi-modality of NFTs, and (3) regard NFTs as both artworks and financial assets. Numerical experiments with real-world NFT transaction data demonstrated that our model outperforms various baseline models while each of our modeling choices was proven to be effective. We believe this work can inspire other researchers to develop more advanced recommender systems for NFTs, ultimately leading to more efficient NFT marketplaces and further flourishing of the relevant industries in the future.

\bibliographystyle{ACM-Reference-Format}
\bibliography{ref_list}

\appendix\label{sec:Appendix}

\section{Data Description} \label{sec: Data Description}
\subsection{Overview}

\begin{table}[!htbp]
\centering

\begin{tabular}{l|c|c|c}
\toprule
Collection & Users & Items & Interactions \\
\hline
%\hline

\midrule

BAYC & 1,230 & 6,726 & 13,737 \\
Cool Cats & 1,357 & 6,824 & 14,890 \\
Doodles & 804 & 4,771 & 7,250 \\
Meebits & 1,184 & 6,693 & 21,104 \\
\hline
Total & 4,575 & 25,014 & 56,981
\end{tabular}
\caption{The number of users, items and interactions for each collection.}
\label{data_description}
\end{table}

\noindent Table \ref{data_description} shows the number of users, items, and interactions for each collection used in our experiment. BAYC, Cool Cats, and Meebits cover a period from September 1, 2021, until March 10, 2023. The transaction data for Doodles cover a period from October 17, 2021, until March 10, 2023, due to their relatively late release.

\subsection{Evaluating Data Density and Transaction Thresholds}

\begin{figure*}[htbp]
\centerline{\includegraphics[width=1\columnwidth]{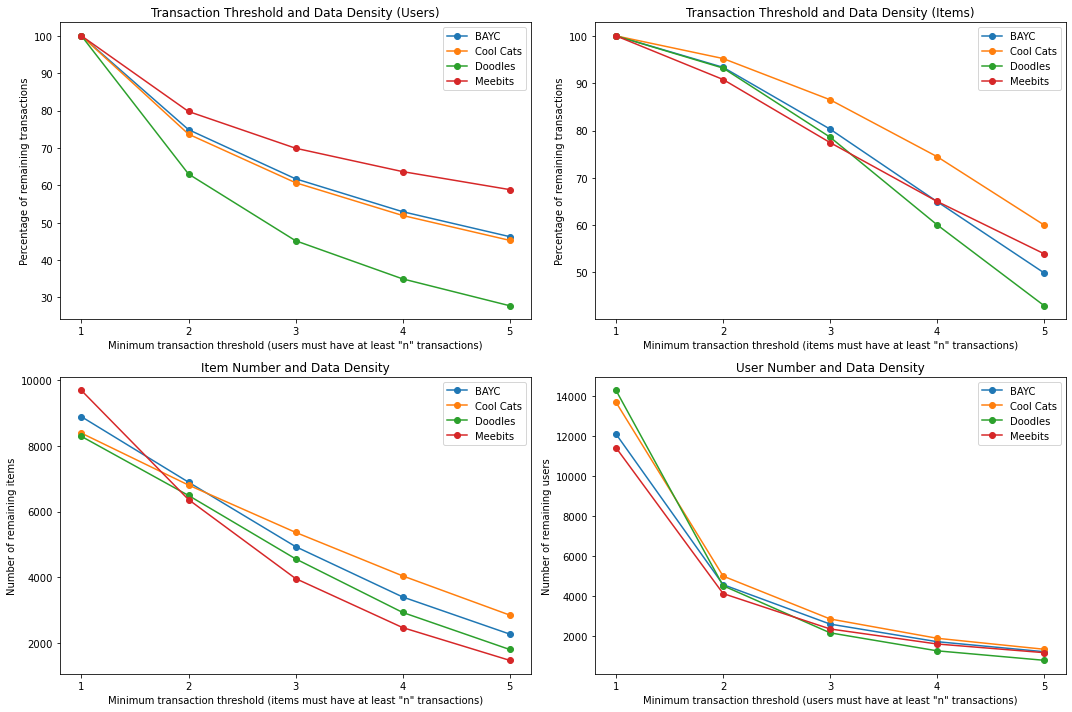}}
\setlength{\abovecaptionskip}{2pt}
\captionsetup{skip=1pt}
\caption{Transaction Threshold and Data Density}
\label{Transaction Threshold and Data Density}
\end{figure*}

One of the critical steps in building an effective recommender system is the evaluation of data density and transaction thresholds in the dataset. To this end, we conducted an analysis using the complete set of transactions. Figure \ref{Transaction Threshold and Data Density} illustrates that a large portion of users and items only made a single appearance in the user-item interaction matrix, and a significant drop in the number of users, items, and interactions was evident upon incrementing the minimum threshold. Based on these observations, we determined that a minimum transaction threshold of 5 is appropriate for the user inclusion criteria. Although no specific exclusion criteria were set for items, it was ensured that only items possessing a full array of features, including text, image, price, and transaction, were considered for inclusion.

\subsection{Power Law Distribution}

\begin{figure*}[htbp]
\centerline{\includegraphics[width=1\columnwidth]{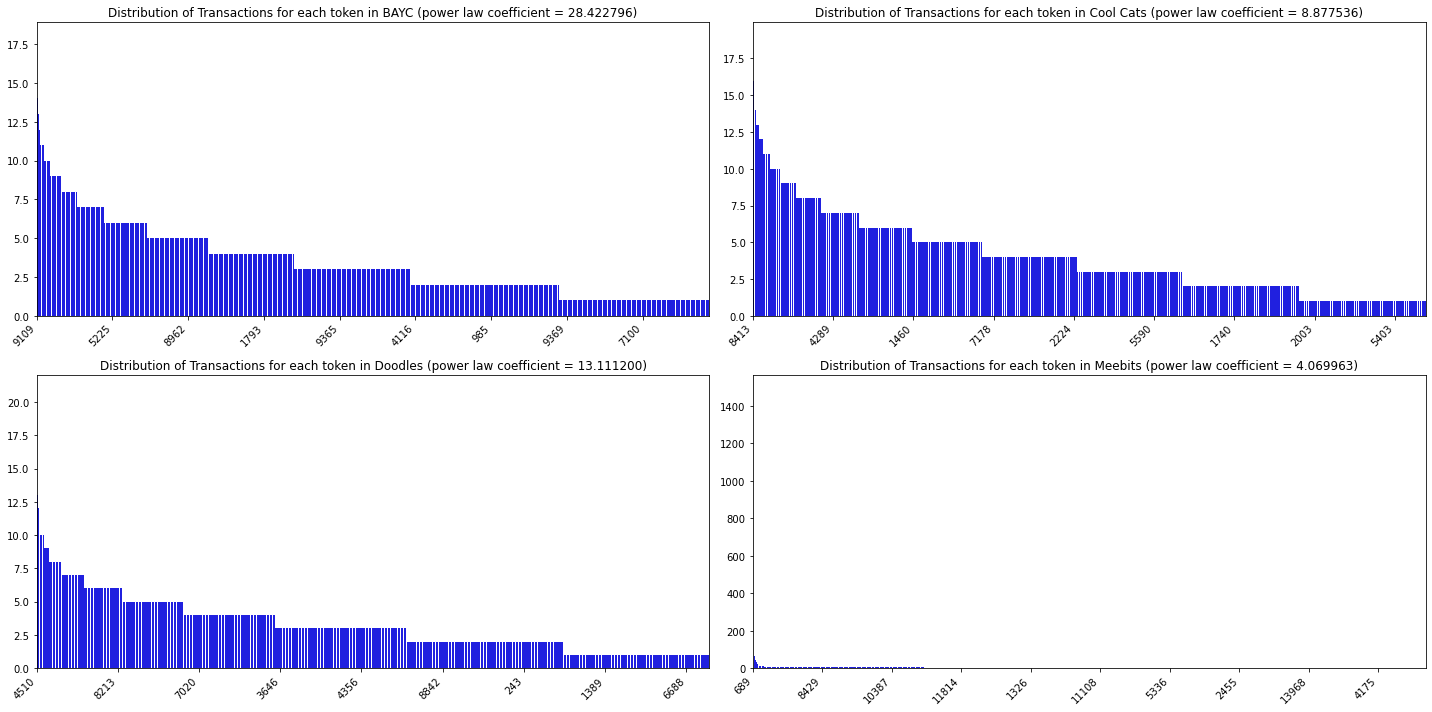}}
\setlength{\abovecaptionskip}{2pt}
\captionsetup{skip=2pt}
\caption{Power Law Distribution of Our Data}
\label{fig:power_law_our_data}
\end{figure*}

\begin{figure*}[!htbp]
\centerline{\includegraphics[width=1\columnwidth]{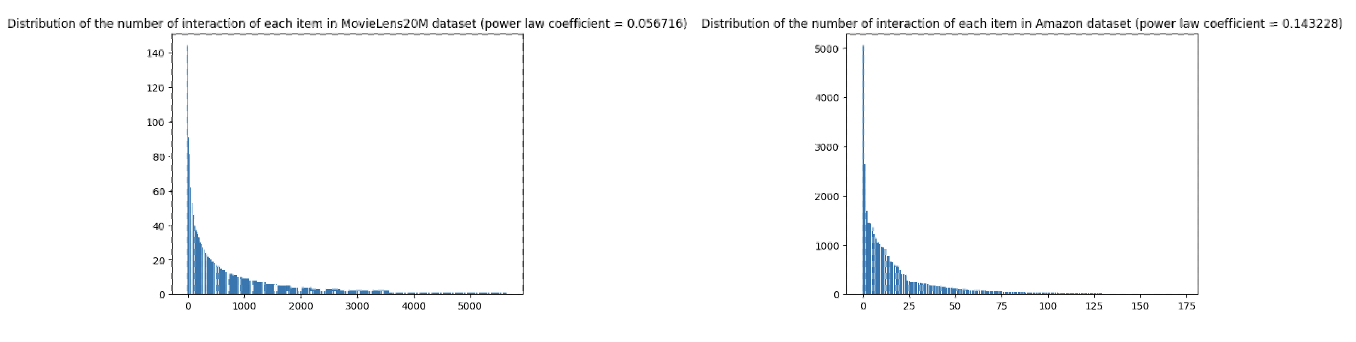}}
\setlength{\abovecaptionskip}{2pt}
\captionsetup{skip=2pt}
\caption{Power Law Distribution of Benchmark Data}
\label{fig:power_law_benchmark}
\end{figure*}

Power law distributions provide insights into the relative popularity and concentration of tokens in a dataset. To gain a deeper understanding of our dataset, we analyze its power law distribution and compare it to a benchmark dataset. This comparison enables us to gauge the alignment of our dataset with standard patterns and to identify any distinctive characteristics. For a fair comparison, we have selected a random sample of 14,000 interactions from the benchmark dataset, which corresponds to the average number of interactions in our dataset.

\noindent The distribution for our datasets in figure \ref{fig:power_law_our_data} shows a weaker power law distribution compared to a benchmark dataset in figure \ref{fig:power_law_benchmark}, suggesting that the popularity of items in our dataset is more evenly distributed. This could potentially make the recommendation task more challenging if relying solely on the collaborative filtering method. In a context where popularity serves as a straightforward and effective indicator for recommendations, it becomes less informative in our dataset due to this more evenly distributed popularity.

\subsection{Token Price Movement Analysis}
In order to incorporate price movement labels for multi-task learning, we have conducted an examination of the price movements for each token. We compute the label by examining the price difference between the current transaction and the one that follows. The label is binary: '1' signifies an expected price increase in the next transaction, while '0' denotes either a price decrease or situations where subsequent transactions don't exist for comparison. The columns in the table represent the following: the number of tokens for which the latest transaction price is greater than the initial transaction price, the number of tokens for which the latest transaction price is greater than the mean of the remaining transaction prices, and the number of tokens for which the mean of the second half of the transaction prices is greater than the mean of the first half of the transaction prices. In the table, each row represents a collection, and the values indicate the percentage of tokens that exhibit the respective price movements.

\begin{table}[htbp]
\begin{tabularx}{\linewidth}{|X|X|X|X|}
\hline
Collection & Latest Price > First Price & Latest Price > Avg. Price (Excluding Last) & Avg. Price (First Half) < Avg. Price (Last Half) \\
\hline
BAYC & 59\% & 61\% & 59\% \\
Cool Cats & 47\% & 47\% & 50\% \\
Doodles & 58\% & 48\% & 59\% \\
Meebits & 45\% & 45\% & 47\% \\
\hline
\end{tabularx}
\caption{Token Price Movement Analyses}
\label{Token Price Movement Analyses}
\end{table}

% \begin{tabular}{|c|c|c|c|}
% \hline
% Token & Latest Price > First Price & Latest Price > Avg. Price (Excluding Last) & Avg. Price (First Half) < Avg. Price (Last Half) \\
% \hline
% BAYC & 59\% & 61\% & 59\% \\
% Cool Cats & 47\% & 47\% & 50\% \\
% Doodles & 58\% & 48\% & 59\% \\
% Meebits & 45\% & 45\% & 47\% \\
% \hline
% \end{tabular}

\section{Side Information Preparation} \label{sec: Side Information Preparation}
\subsection{Item Features}
\begin{itemize}
    \item \textbf{Image}
    We employ Convolutional Auto-Encoder (CAE) to get representations from the NFT images. We standardize all images to a shape of 128 * 128 * 3, where 3 represents the RGB color spectrum. CAE model consists of an encoder and a decoder, both comprising of eight fully connected layers. The encoder utilizes a 33 convolutional kernel and 22 max pooling, while the decode employs a 33 convolutional kernel along with 22 upsampling. All non-linear functions in the model are implemented using the ReLU activation function. The model is trained for 100 epochs, with the objective of minimizing the Mean Squared Error (MSE) loss. The final image embeddings are obtained by employing only the encoder of the CAE, which reduces the size down to 8 * 8 * 1. After flattening the output, we receive a 64-dimension representation for each image. This refined data is then ready for subsequent modeling stages.
    \item \textbf{Text}
    The text data for each item is comprised of discrete words each describing each of the visual properties like ‘Background color’, ‘body’, ‘outfit’, and ‘hair’. Items within the same collection share the same types of visual properties whereas they tend to vary across collections. Cool cats, for example, is a collection of blue cat avatars with each artwork possessing a unique body, hat, face, and outfit, whereas Bored Ape Yacht Club is a collection of monkey avatars having a slightly different types of properties like ‘fur’. Among all, we only have considered six types of properties with the fewest missing values for each collection apart from Cool cats, for which we considered all available 5 types of properties for generating item embeddings. We then processed each descriptive word into a 300-dimension word embedding by fetching the corresponding embeddings from a pre-trained Word2Vec model. If a particular word was not found in this model, we filled it with zero padding. It's worth noting that while a majority of visual attributes were described by a single word, those composed of multiple words, like 'short red hair' for 'Hair', we used the sum of each word's embeddings instead. Each word embedding was then concatenated with other embeddings related to the same item, hence, each item's word embedding size ranged from 1500 to 1800, depending on the number of visual traits considered.
    \item \textbf{Price, Transaction}
    Unlike content features, obtained representations from transaction features do not have sufficient embedding size since there is one average value per each item. So we duplicated the same value to have 64 dimensions.
\end{itemize}

\subsection{Price Movement Label}
We generate price movement label to represent the change in the price of a token between two transactions. The purpose of this feature is to classify whether the price of a token is going to increase or not in the next transaction. We calculate the price difference of each token between the current and subsequent transaction and label it 1 for upward movement, 0 for downward movement or any instances where no subsequent transactions are available for comparison. This kind of information can be valuable in predicting user behavior, as users may behave differently based on whether they anticipate a price increase.

\section{Hyperparameter Details} \label{sec: Hyperparameter}
We optimized \texttt{NFT-MARS} using the Adam optimizer, a learning rate of 0.01, and 50 epochs. Hyperparameter tuning involved a random search using Recall@50 as an indicator, with search ranges including the dimensions ($d$) of the graph’s final node representation [128, 512], loss alpha ($\alpha$) [0.1, 0.2], batch size [1024, 4096], number of hops ($L$) [1, 2, 3], and regularization weight [0.1, 0.001]. Best hyperparameter values for \texttt{NFT-MARS} are specified in below table \ref{tab:MARS hyperparameter}.

\begin{table}[h]
    \centering
    \begin{tabular}{|l|l|l|l|l|l|l|}
        \hline
        collection & seed & dimension ($d$) & loss alpha ($\alpha$) & batch size & number of hops ($L$) & regularization weight \\
        \hline
        BAYC       & 2023 & 128            & 0.2                   & 1024       & 2                     & 0.1                   \\
        Coolcats   & 2024 & 512            & 0.2                   & 1024       & 1                     & 0.1                   \\
        Doodles    & 2022 & 512            & 0.1                   & 1024       & 3                     & 0.001                 \\
        Meebits    & 2022 & 512            & 0.1                   & 1024       & 1                     & 0.001                 \\
        \hline
    \end{tabular}
    \caption{Optimal Hyperparaemter Details for \texttt{NFT-MARS}}
    \label{tab:MARS hyperparameter}
\end{table}

\noindent Same optimizer, Adam, was also used for optimization of MGAT model. As for tuning hyperparameters for MGAT, we fix the loss alpha ($\alpha$) to 0 test and compare the effectiveness of the multi-task learning, and learning rate to 0.01. We then select the dimensions ($d$) of the graph’s final node representation from [128, 512], batch size from [1024, 4096], regularization weight from [0.1, 0.001], and number of hops ($L$) from [1, 2, 3]. Best hyperparameter values for MGAT are specified in below table \ref{tab:MGAT hyperparameter}.

\begin{table}[h]
    \centering
    \begin{tabular}{|l|l|l|l|l|l|}
        \hline
        collection & seed & dimension ($d$) & batch size & number of hops ($L$) & regularization weight \\
        \hline
        BAYC       & 2022 & 128            & 4096       & 1                    & 0.1                   \\
        Coolcats   & 2024 & 512            & 1024       & 1                    & 0.001                 \\
        Doodles    & 2023 & 512            & 1024       & 1                    & 0.001                 \\
        Meebits    & 2024 & 128            & 1024       & 1                    & 0.001                 \\
        \hline
    \end{tabular}
    \caption{Optimal Hyperparameter Details for MGAT}
    \label{tab:MGAT hyperparameter}
\end{table}

\noindent Baseline models’ hyperparameters were also tuned, in regards to embedding size, learning rate, and dropout ratio. Specific details regarding the optimal hyperparameter values, hyperparameter search range can be found in \url{https://anonymous.4open.science/r/RecSys2023-93ED}.

\end{document}